\documentclass[letterpaper, 10 pt, conference]{ieeeconf}

\IEEEoverridecommandlockouts
\usepackage{cite}
\usepackage{times}
\usepackage[pdftex]{graphicx}
\graphicspath{{../Figures}}

\DeclareGraphicsExtensions{.pdf,.jpeg,.png,.jpg}
\usepackage{caption}
\usepackage[normalem]{ulem}
\usepackage{subcaption}
\usepackage{epstopdf}
\usepackage[cmex10]{amsmath}
\usepackage{amssymb}
\usepackage{amsfonts}
\usepackage{mathrsfs}
\usepackage{mathabx}
\usepackage{algorithmic}
\usepackage{array}
\usepackage{mdwmath}
\usepackage{mdwtab}
\usepackage{stfloats}
\usepackage{fancybox,color}
\usepackage[lined,commentsnumbered]{algorithm2e}
\usepackage{url}
\usepackage[utf8]{inputenc}
\usepackage[nodisplayskipstretch]{setspace}
\usepackage{hyperref}
\usepackage{ulem}
\hypersetup{
    colorlinks=true,
    linkcolor=blue,
    filecolor=blue,      
    urlcolor=blue,
}
\usepackage{lipsum}

\def\ie{{\it i.e.,\ }}

\newtheorem{theorem}{Theorem}

\newtheorem{definition}[theorem]{Definition}

\usepackage{ulem}

\title{\LARGE \bf
Differential Geometric Approach to Trajectory Planning:\\
Cooperative Transport by a Team of Autonomous Marine Vehicles 
}

\author{Hadi Hajieghrary, Dhanushka Kularatne $^{1}$, and M. Ani Hsieh $^{2}$
\thanks{$^{1}$ Emails: hh449@drexel.edu. The author is with the SAS Laboratory, Mechanical Engineering \& Mechanics Department, Drexel University, Philadelphia, PA 19104, USA.}%
\thanks{$^{2}$ Email:  \{dkul, m.hsieh\}@seas.upenn.edu. Dhanushka Kularatne and  M. Ani Hsieh are with the GRASP Laboratory, University of Pennsylvania, Philadelphia, PA 19104, USA.}%
\thanks{We gratefully acknowledge the support of the National Science Foundation (NSF) grant CMMI-1462825.}%
}

\begin{document}
\maketitle
\thispagestyle{empty}
\pagestyle{empty}

\begin{abstract}
In this paper we addressed the cooperative transport problem for a team of autonomous surface vehicles (ASVs) towing a single buoyant load. We consider the dynamics of the constrained system and decompose the cooperative transport problem into a collection of subproblems.  Each subproblem consists of an ASV and load pair where each ASV is attached to the load at the same point.  Since the system states evolve on a smooth manifold, we use the tools from differential geometry to model the holonomic constraint arising from the cooperative transport problem and the non-holonomic constraints arising from the ASV dynamics. We then synthesize distributed feedback control strategies using the proposed mathematical modeling
framework to enable the team transport the load on a desired
trajectory. We experimentally validate the proposed strategy using a team of micro ASVs.
\end{abstract}

\section{INTRODUCTION}
We are interested in cooperative manipulation strategies that enable a team of robots to transport large objects that cannot be transported by single robots alone. Specifically, we are interested in cooperative manipulation and transportation of objects in marine and littoral environment when the object is attached to the robots via a cable. In these systems, the holonomic constraints arise from the need to coordinate with other members of the team and are imposed on agents' states. The non-holonomic constraints arise from the agents' dynamics.  This is a highly constrained system that can benefit from an optimal control solution that is scalable and amenable to changes in the team size.  Tabuada et al. in \cite{Tabuada2005} showed how it is possible to design trajectories and synthesize controllers for a group of agents that simultaneously satisfy the dynamics of each agent and a set of holonomic constraints.  This problem is called the formation control abstraction \cite{Belta2004}.


The problem of transporting a cable suspended load has been extensively studied in the robotics domains \cite{Michael2011,Wu2014,Lee2014,Cruz2015,Zeiaee2016}. The kinematic problem was considered in for a team of aerial vehicles \cite{Michael2011}.  Cooperative transport of a heavy load by a team of quadrotors was accomplished using geometric feedback control in \cite{Wu2014}.  A similar approach was used in \cite{Lee2014}, however in this work the control action was decomposed into parallel and normal components to the vehicle trajectories.  A hybrid control strategy was used in \cite{Cruz2015} to enable the vehicle to execute lift maneuvers while minimizing the load swing.  The cooperative load carry problem for mobile grounds robots subject to hard input constraint was discussed in \cite{Zeiaee2016}.  

Since the cooperative transport problem is a specific instantiation of a mechanical system with constraints, the system can be analyzed using the techniques presented in \cite{Lewis2000}.  Similar control theoretic strategies which analyzes controllability and motion planning for multi-body systems with non-holonomic constraints include \cite{Shen2003,Cortes2002,Colombo2017}. The optimal trajectory planning problem can be posed as finding a geodesic path on the (sub)Reimannian manifold given by the system's configuration space is presented in \cite{Stefani2014}. Such geometric approaches provide a well-suited tool set for designing optimal control policies for these systems \cite{Montgomery2006}.  The extension of these geometric approaches to mechanical systems subject to non-holonomic constraints is discussed in \cite{Bloch2015}. 

Nevertheless, cooperative transport of a cable towed cargo by a team of autonomous surface vehicles (ASVs) introduce new challenges.  Similar to their non-marine counterparts, these vehicles are subject to limitations on their maneuverability.  Vehicle and load inertial effects become more significant in the marine environment.  In this work, we build upon our existing work \cite{Hajieghrary2017} and present a solution to the control abstraction problem for cooperative load transport by a team of ASVs. Different from our previous work, we explicitly taking into account the vehicle dynamics in this work. Rather than computing a set of trajectories for the vehicles and load, we synthesize a set of control inputs for each vehicle to enable the team to maintain the desired formation as they transport the load along the desired trajectory.  We show the feasibility of the approach in experiment.  Our results suggest that the proposed strategy is robust in the presence of disturbances and model uncertainties.

The rest of this paper is organized as follows: We formulate our problem and present our modeling framework in Section \ref{Sec: PROBLEM FORMULATION}. Section \ref{Sec: COOPERATIVE CONTOLLER DESIGN} presents our methodology. The experimental results are presented in Section \ref{Sec:RESULTS}.  The paper concludes with a summary of our contribution and discussion on future work in Section \ref{Sec:CONCLUSION}.

\section{PROBLEM FORMULATION}
\label{Sec: PROBLEM FORMULATION}

\subsection{Preliminaries}
\label{SubSec:Preliminaries}


We begin with some preliminaries and adopt the notation presented in \cite{Bullo2004}.  The trajectory of a system is a curve on the topological manifold of the system states, $\gamma(t):\mathbb{R}\to Q$. The velocity along this curve at any point $\gamma (0) = p \in Q$, is a linear map from the vector space of the smooth functions on the manifold to the real numbers, and is defined as
\begin{eqnarray}
 C^{\infty}(Q) \ni f \mapsto \mathit{v}_{p}(f)=(f \circ \gamma(0))^\prime \in \mathbb{R}. 
\end{eqnarray} 
where the operator $\circ$ indicates function composition. At each point of the manifold, the set of the velocities for all trajectories passing through that point constitute a vector space called the \textit{Tangent Vector Space} of the manifold $Q$ at the point $p$, {\it i.e.}, $T_pQ$. This set is a vector space equipped with addition and scalar multiplication \cite{Bullo2004}.

Let $\big(\mathcal{U} \subset Q, \mathit{X}(\mathcal{U}) \in \mathbb{R}^n\big)$ be a chart on the smooth manifold $Q$. A tangent vector, $\mathit{v}_{\gamma(t),p}$, is an operator which acts on a function and returns a real number, and is given by
\begin{align}
	\mathit{v}_{\gamma(t),p}(f)&= (f \circ \gamma)^\prime (0) = (f \circ \mathit{X}^{-1} \circ \mathit{X} \circ \gamma)^\prime (0)\\
	&= {(\mathit{x}^i \circ \gamma)}^\prime (0) \cdot \partial_i(f \circ \mathit{X}^{-1}) (\mathit{X}(p)) \nonumber
\end{align}
where we employ Einstein's summation notation.  We note that the second part of the tangent vector is independent of the curve $\gamma (t)$ and thus an independent entity on the manifold. The representation of these objects in the chosen chart is called the chart-induced basis of the tangent space, $\{(\frac{\partial}{\partial x^i})_{p}\}$. We note that the partial differentials in $\frac{\partial}{\partial x^i}(f)$ are symbols and not operators since $f$ is not a function of $x_i$s but instead a function of the points on $Q$. The set of these operators in a chart forms a basis for the tangent space of the manifold $Q$ at the point $p$, \ie $T_pQ$.

The set of all linear maps from the tangent space to the real numbers equipped with the proper addition and scalar multiplication operators forms a vector space called the \textit{cotangent space}. For example, the gradient of a function, $df$, is a covector, defined as $T_{p}Q \ni \mathcal{\chi} \mapsto (df)_{p}(\mathcal{\chi})=\mathcal{\chi}_{p}(f) \in \mathbb{R}$.
And, like any other vector, a co-vector can also be written in components with respect to a chart, $\big(\mathcal{U}, \mathit{X}\big)$:
\begin{eqnarray}
	[(df)_{p}]_i=(df)_{p}\big(\frac{\partial}{\partial x^i}\big)=\partial_i\big(f \circ x^{-1}\big)\big(x(p)\big)
\end{eqnarray}
The components of the chart themselves are maps from a subset of $Q$ to the real numbers, $x^{i}: \mathcal{U} \in Q \to \mathbb{R}$. The set of co-vectors defined using these functions constitutes a basis for the cotangent space, $ \bigg\{\big(dx^1\big)_p,\big(dx^2\big)_p,\ldots,\big(dx^d\big)_p\bigg\}$. These are the dual basis of the tangent basis, $\big(dx^i\big)_p\big(\frac{\partial}{\partial x^j}\big) = \frac{\partial x_i}{\partial x^j} = \delta_j^i$, where, $\delta_j^i$ is Dirac's delta function. 

Vector fields are used to calculate the directional derivative of a function defined on a manifold. In differential geometry, \textit{connections} (linear, covariant, or affine connection),  $\nabla_{\mathcal{\chi}}$, are operators with a predefined list of properties.
\begin{definition}
	An (affine) connection on a smooth manifold is a map which takes a pair consisting of a vector field, $\mathcal{\chi}$, and a (p,q)-tensor field, $T$ and $S$, and returns a (p,q)-tensor field such that it satisfies the following axioms: 
	\begin{enumerate}
		\item $\nabla_{\mathcal{\chi}} f=X(f)$, for $f \in C^{\infty}(Q)$;
        \item $\nabla_{\mathcal{\chi}} (T+S) = \nabla_{\mathcal{\chi}} T + \nabla_{\mathcal{\chi}} S$;
        \item (Leibniz rule) $\nabla_{\mathcal{\chi}} \big(T(\omega,\mathit{\psi})\big) = \big(\nabla_{\mathcal{\chi}}T\big)(\omega,\mathit{\psi}) + T(\nabla_{\mathcal{\chi}}\omega,\mathit{\psi}) + T(\omega,\nabla_{\mathcal{\chi}}\mathit{\psi})$;
	\end{enumerate} 
\end{definition}
This definition of the \textit{connection} is complete when the  list of predefined properties results in a uniquely defined geometric operator.  However, to turn this object into a chart, one needs an additional structure on the manifold.  Since application of the operator on 
the basis of the tangent vector space, $\{\frac{\partial}{\partial x^i}\}$ and should result in a tangent vector field that is spanned by the tangent space at that point, this results in
\begin{eqnarray}
	\nabla_{(\frac{\partial}{\partial x^i})_p}(\frac{\partial}{\partial x^j})_p = \Gamma^{k}_{\{X\}ji} (\frac{\partial}{\partial x^k})_p.  
\end{eqnarray}
We note that coefficients in the above expression are ${(dim(Q))}^3$ and may be defined in many ways, $\Gamma^{k}_{\{X\}ji}:\mathcal{U} \in Q \to \mathbb{R}$. The choice of these functions fixes the action of $\nabla$ on \textit{any} tensor field defined on the manifold.

A vector field, $\mathit{\chi}$, is called \textit{Parallel Transported} along an smooth curve, $\gamma:\mathbb{R} \to Q$, if $\nabla_{\mathit{v}_{\gamma(t)}}\mathit{\chi}\equiv0$. The curve $\gamma(t)$ is said to be {\it Auto-Parallel Transported} if $\nabla_{\mathit{v}_{\gamma(t)}}\mathit{v}_{\gamma(t)}\equiv0$. This expression is written using a chart as 
\begin{eqnarray}
{\ddot{\gamma}}^{i}_{\{X\}}+\Gamma^{i}_{\{X\}jk}{\dot{\gamma}}^{j}_{\{X\}}{\dot{\gamma}}^{k}_{\{X\}}=0
\label{eq:Auto-Parallel}
\end{eqnarray}

A metric $G$ on a smooth manifold is a (0,2)-tensor field satisfying 1) Symmetry: $G(\mathit{\chi},\psi)=G(\psi,\mathit{\chi}), \forall \mathit{\chi},\psi \in \Gamma^{\infty}(TM)$, and 2) Non-degeneracy: $G(\mathit{\chi},\psi)=0 \Leftrightarrow \big(\mathit{\chi}=0~,~\psi=0\big)$. The length of a smooth curve is a real number $L_G[\gamma]=\int_a^b \sqrt{G(\mathit{v}_{\gamma(t)},\mathit{v}_{\gamma(t)})}dt$. In a torsion free connection the necessary and sufficient conditions for an auto-parallel transported curve, \ie the straightest curve, to be the shortest one, measured by the metric $G$, is $(\nabla G)\equiv 0$, called the \textit{geodesic}.

Consider the length of the trajectory as the cost functional. The following differential equation is the Euler-Lagrange equation written in the components of the curve $\gamma (t)$:
\begin{eqnarray}
	\ddot{\gamma}^{q}+(G^{-1})^{qm}\Big( \frac{\partial G_{mj}}{\partial x^i} + \frac{\partial G_{mi}}{\partial x^j} + \frac{\partial G_{ij}}{\partial x^m} \Big) \dot{\gamma}^{i} \dot{\gamma}^{j} = 0,
    \label{Euler-Lagrange equation}
\end{eqnarray}
where $G^{-1}$ is defined as the inverse metric with components satisfying $(G^{-1})^{qm}(G^{-1})_{mj}=\delta^q_j$ \cite{Bullo2004}. There is an obvious similarity between the the Euler-Lagrange conditions in \eqref{Euler-Lagrange equation} and the auto-parallel transported curve written in its components, as in \eqref{eq:Auto-Parallel}. By defining the connection coefficient functions as: \begin{eqnarray}
	\Gamma^{q}_{ij} \triangleq  (G^{-1})_{qm} \Big( \frac{\partial G_{mj}}{\partial x^i} + \frac{\partial G_{mi}}{\partial x^j} + \frac{\partial G_{ij}}{\partial x^m} \Big)
\end{eqnarray}
the auto-parallel transported curve will also be the geodesic of the metric $G$. Such a torsion free connection is called the Levi-Civita Connection, and the connection coefficient functions for Levi-Civita connection are called Christoffel symbols. In a manifold defined with these metric and connection coefficients the straightest path will be the shortest one \cite{doCarmo2013}.  

\subsection{Problem Statement}\label{SubSec:Problem Statement}

\begin{figure}
	\centering
	\includegraphics[width=0.7\linewidth]{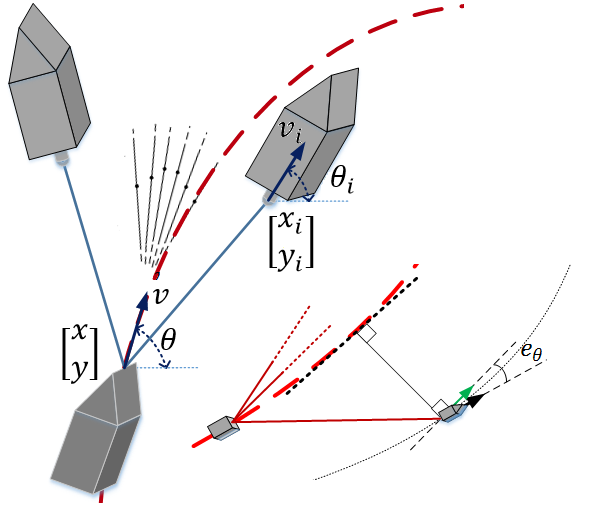}
	\caption{Schematic of cooperative transport of a buoyant load by two autonomous vehicles. and attitude control objective.}
    \vspace{-6mm}
	\label{fig: transport Schematic}
\end{figure}
Consider a team of $N$ Autonomous Surface Vehicles (ASVs) which are transporting a load along a desired reference trajectory, as shown in Fig. \ref{fig: transport Schematic}. The load is connected to each vehicle by a flexible inextensible cable. Let $\mathbf{q}_i = [x_i, \, y_i, \, \theta_i]^T$ denotes the pose of the $i^{th}$ ASV and $\mathbf{q}_L = [x, \, y, \, \theta]^T$ be the pose of the load. The $i^{th}$ vehicle's and the load's configuration space together define a smooth manifold, $\mathcal{Q}$, for which a representative chart is given by these state variables. The kinematics of the vehicle and the load constitute non-holonomic constraints on this manifold. These constraints can be expressed in terms of the derivatives of the components of the chart map:
\begin{eqnarray}
	\begin{cases}
		\dot{x}=v \cos \theta\\
		\dot{y}=v \sin \theta\\
		\dot{\theta}=\omega
	\end{cases},
    \qquad \qquad
	\begin{cases}
		\dot{x}_i=v_i \cos \theta_i\\
		\dot{y}_i=v_i \sin \theta_i\\
		\dot{\theta}_i=\omega_i
	\end{cases}.
	\label{eq:UniCycle_Non-holonomic_Constraints}
\end{eqnarray} 
We note \eqref{eq:UniCycle_Non-holonomic_Constraints} are simply the kinematics of the $i^{th}$ vehicle and the load.  It also describes the non-holonomic constraints for the system and are the generators of a distribution on the tangent space of the configuration manifold, $T\mathcal{Q}$ \cite{Bullo2004}: 
\begin{align}
	\Delta = span \Big\{ e_1 &= \cos \theta \frac{\partial}{\partial x} + \sin \theta \frac{\partial}{\partial y},\quad e_2 =\frac{\partial}{\partial \theta},\\
    e_3 &=\cos \theta_i \frac{\partial}{\partial x^i} + \sin \theta_i \frac{\partial}{\partial y^i},\quad e_4 = \frac{\partial}{\partial \theta^i}\Big\}.\nonumber
    \label{Non-Holonomic_Distribution}
\end{align}
Every tangent vector field along any trajectory of the system can be written as a linear combination of the generators of the distribution, \ie $\mathit{v}_{\gamma(t)} = v.e_1 + \omega. e_2 + v_i.e_3 + \omega_i. e_4$. 

There is a unique co-distribution which annihilates this distribution, $\alpha_i(e_j) \equiv 0$ for any $\alpha_i \in \Lambda$. The number of linearly independent co-vectors in this co-distribution is equal to the dimension of the manifold, $n=6$, minus the number of the generators of the distribution, $m=4$. One set of generators for this co-distribution is calculated as:
\begin{eqnarray}
	\Lambda =span\{
		\sin \theta_i dx_i - \cos \theta_i dy_i,~ \sin \theta dx - \cos \theta dy\}.
	\label{Co-distribution_Lambda}
\end{eqnarray}

Cooperation among the vehicles to transport the load imposes a set of holonomic constraints on the configuration variables of each vehicle an the load given by 
\begin{eqnarray}
	\mathit{C}_i:~{\big(x-x_i\big)}^2+{\big(y-y_i\big)}^2 = l_i^2
	\label{eq:holonomic}
\end{eqnarray}
for all $i = 1, \ldots, N$, where $l_i$ denotes the length of the cable connecting the $i^{th}$ vehicle to the load. The image of the map $\mathit{C}_i$ is a sub-manifold of $\mathcal{Q}$ and the differential of this map is a co-distribution which annihilates all vectors in the tangent space of this sub-manifold, $d\mathit{C}_i=\Big\{\big(x-x_i\big)(dx-dx_i) + \big(y-y_i\big)(dy-dy_i)\Big\}$. The unified single co-distribution $\Sigma = d\mathit{C}_i \oplus \Lambda$ annihilates the part of the distribution $\Delta$ which also satisfies the holonomic constraint \cite{Hajieghrary2017}. This co-distribution has constant algebraic rank of three which guarantees the existence of a distribution that will be globally annihilated by it \cite{Bullo2004}. The components of this distribution are found as $\mathcal{D} = span\Big\{\frac{\partial}{\partial \theta^i}, \,\quad \frac{\partial}{\partial \theta}, \quad h_i\cos \theta  \frac{\partial}{\partial x^i} + h_i \sin \theta \frac{\partial}{\partial y^i} + h \cos \theta_i \frac{\partial}{\partial x} + h \sin \theta_i \frac{\partial}{\partial y}\Big\}$, where $h\triangleq \big(x-x_i\big) \cos \theta + \big(y-y_i\big) \sin \theta$ and $h_i\triangleq \big(x-x_i\big) \cos \theta_i + \big(y-y_i\big) \sin \theta_i$. The tangent vector field along any trajectory of the system \eqref{eq:UniCycle_Non-holonomic_Constraints} that satisfies the constraints \eqref{eq:holonomic} is a linear combination of the vectors in $\mathcal{D}$.

In \cite{Hajieghrary2017} we considered the kinematics of this problem. Here, we incorporate the dynamics of the vehicles in the synthesis of feedback controllers that enable them to cooperatively transport the load along the desired reference trajectory.  We do this by employing an Euler-Lagrange approach and use the Lagrange-d'Alembert principle to include the control forces in the system.

\section{TRAJECTORY DESIGN FOR COOPERATIVE TRANSPORT}
\label{Sec: COOPERATIVE CONTOLLER DESIGN}

\subsection{Kinematics of the Cooperative Action}
\label{SubSec: Kinematics of the Cooperative Action}

The distribution $\mathcal{D}$ consists of three tangent vectors. However, the tangent space of the configuration manifold, $Q$, is of dimension six. We need to add three extra tangent vector fields, $X_4 = \cos \theta \frac{\partial}{\partial x} + \sin \theta \frac{\partial}{\partial y}$, $X_5 = \sin \theta \frac{\partial}{\partial x} - \cos \theta \frac{\partial}{\partial y}$, and $X_6 = \sin \theta_i \frac{\partial}{\partial x^i} - \cos \theta_i \frac{\partial}{\partial y^i}$, to the set of basis of $\mathcal{D}$ to achieve an involutive distribution and consequently an integrable one.

The integral manifold of this integrable distribution is an immerse manifold in the base manifold of the distribution. The generators of this distribution constitute a set of basis for the tangent space of this immerse manifold. In the context of this work, these two manifolds are the same. A similar approach can be taken about the distribution $\Delta$. Whereby, adding two extra tangent vectors, $e_5= \sin \theta \frac{\partial}{\partial x} - \cos \theta \frac{\partial}{\partial y}$ and $e_6= \sin \theta_i \frac{\partial}{\partial x^i} - \cos \theta_i \frac{\partial}{\partial y^i}$, turns it to an involutive one. Again, the set $\{e_i\}$ constitutes another alternative set of basis for the tangent space of the configuration manifold.

Any tangent vector field in $TQ$, including the tangent vectors along the trajectory of the system, can be written in these three bases, namely, the chart-induced basis for the tangent space of the state space, $\big\{\frac{\partial}{\partial x}~,~\frac{\partial}{\partial y}~,~\frac{\partial}{\partial \theta}~,~\frac{\partial}{\partial x^i}~,~\frac{\partial}{\partial y^i}~,~\frac{\partial}{\partial \theta^i}\big\}$, the involutive set of vector fields that constitute the kinematics, $\big\{e_i\big\}_{i=1,2...,6}$, and the involutive set of vector fields of the kinematics of the constrained system, $\big\{X_i\big\}_{i=1,2...,6}$. 

We are interested in a trajectory for which the tangent space is restricted to the distribution generated by $X_1$, $X_2$, and $X_3$. Such a trajectory satisfies both the non-holonomic constraints resulting from the system kinematics and the holonomic constraint posed by the cooperative transport of the load.  A projection map can be constructed from the total tangent space of the configuration manifold, $TQ$, to a subspace that is spanned by these vector fields, \ie $X_1$, $X_2$, and $X_3$. The components of such a map in the basis $\big\{X_i\big\}$ can conveniently be presented by 
\begin{eqnarray*}
[P]_{X_i}=\begin{bmatrix}
\mathbf{I}_{3 \times 3} & \mathbf{0}\\
\mathbf{0}    & \mathbf{0} 
\end{bmatrix}.
\end{eqnarray*}
For the trajectory generation problem, we define the objective functional to be the kinematic energy of the system. The corresponding Riemannian metrics, $G$, is the differential change in the kinetic energy of the system, and is defined as 
\begin{align}
G=&M(dx \otimes dx + dy \otimes dy) + Jd\theta \otimes d\theta\\
&+ m_i(dx_i \otimes dx_i + dy_i \otimes dy_i) + J_i d\theta_i \otimes d\theta_i.\nonumber
\end{align}
The variables $M$, $J$, $m_i$, and $J_i$ are the masses and the inertias of the load and the $i^{th}$ ASV respectively. The trajectory which minimizes this objective functional is the geodesic of the Riemannian connection. The complete dynamics of the constrained system can be written as: 
\begin{equation}
  \begin{cases}
	{\stackrel{G}{\nabla}}_{\mathit{v}_{\gamma(t)}}\mathit{v}_{\gamma(t)}= P(Y(t))\\
P^{\prime}(\mathit{v}_{\gamma(t)})=0
\label{Constrained Dynamics}
  \end{cases},
\end{equation}
where $P^{\prime}$ is the complementary map of $P$ \cite{Martinez2007,ACCMaithripala2008}.  Using the basis $\{X_i\}$, $P^{\prime}$ can be written as $\big[P^{\prime}\big]_{X_i}=\big[I\big]_{X_i}-\big[P\big]_{X_i}$. The right hand side of the first equation, $P(Y(t))$, is the consequence of d'Alembert's principle which will be discussed in the next section. To constrain the tangent space of such a trajectory to the distribution $\mathcal{D}$, we amend this connection using the compliment of the map $P$ \cite{Lewis2000,Bullo2004}. The resulting connection is called the constrained connection:
\begin{equation}
{\stackrel{\mathcal{D}}{\nabla}}_{\mathit{v}_{\gamma(t)}}\mathit{v}_{\gamma(t)} = {\stackrel{G}{\nabla}}_{\mathit{v}_{\gamma(t)}}\mathit{v}_{\gamma(t)} + ({\stackrel{G}{\nabla}}_{\mathit{v}_{\gamma(t)}}P^{\prime})(\mathit{v}_{\gamma(t)}).
\label{Constrained Geodesic}
\end{equation}
The components of the covariant derivative of the map $P^{\prime}$ with respect to a tangent vector field, \ie $Y=Y^i\frac{\partial}{\partial q^i}$, in the chart induced coordinate frame are
\begin{equation}
\big[{\stackrel{G}{\nabla}}_{Y}P^{\prime}\big]_i^j =\Big ( \frac{\partial {P^{\prime}}_i^j}{\partial q^k}+ {\stackrel{G}{\Gamma^j}}_{(q)kr} {P^{\prime}}_r^i - {\stackrel{G}{\Gamma^r}}_{(q)ki} {P^{\prime}}_r^j \Big ) Y^k,
\label{Covariant Derivative of a Map}
\end{equation}
where ${P^{\prime}}_i^j=dq^i\big({P^{\prime}}(\frac{\partial}{\partial q^j})\big)$. Since the components of the metric tensor $G$ is constant, then ${\stackrel{G}{\Gamma^k}}_{(q)ij}=0$, for any $i,j,k\in{1,2,...,6}$. Therefore, it is easier to write down the components of the covariant derivative of this map in the chart-induced coordinate frame. To do so, we first rewrite the component matrix of this map, $P^{\prime}$, with respect to the chart-induced basis, converted from $\big\{X_i\big\}$ in which the map originally introduced as $[P]_{X_i}$; then, rewrite it in the basis $\{e_i\}$ which represents the kinematics of the system. Let $\textbf{X}=\mathcal{R}\mathbf{\partial_q}$ be the relationship between the components of a tangent vector field written in these two different basis:
\begin{equation}
\mathcal{R} =
\resizebox{0.78\linewidth}{!}
{$
\begin{bmatrix}
h\cos \theta_i & h\sin \theta_i & 0			& h_i\cos \theta & h_i\sin \theta & 0			\\
0			 & 0			& 1			& 0				   & 0				  & 0			\\
0			 & 0			& 0			& 0				   & 0				  & 1			\\
\cos \theta	 & \sin \theta	& 0			& 0				   & 0				  & 0			\\
\sin \theta	 & -\cos \theta	& 0			& 0				   & 0				  & 0			\\
0			 & 0			& 0			& \sin \theta_i	   & -\cos \theta_i	  &	0
\end{bmatrix}
$}.
\end{equation}
The map ${P}^{\prime}$ expressed in the chart induced basis can be written in matrix form as $\big[P^{\prime}\big]_{\partial_q}=\mathcal{R}^{-1}\big[P^{\prime}\big]_{X}\mathcal{R}$. To calculate the coefficients of the constrained connection ${\stackrel{\mathcal{D}}{\nabla}}_{\mathit{v}_{\gamma(t)}}\mathit{v}_{\gamma(t)}$ in the chart induced coordinates, we substitute the derivatives of the components of the $\big[P^{\prime}\big]_{\partial_q}$ into \eqref{Covariant Derivative of a Map}.

The control variables for the system given by  \eqref{eq:UniCycle_Non-holonomic_Constraints} are the coefficients of the generators of the distribution $\Delta$. The distribution $\Delta$ is the subspace of $\mathcal{D}$. We define the generalized Christoffel symbols given by ${\stackrel{\mathcal{D}}{\Gamma^i}}_{(x)jk}$,  as $\stackrel{\mathcal{D}}{\nabla}_{e_k}e_j={\stackrel{\mathcal{D}}{\Gamma^i}}_{(e)jk}e_i$. The tangent vector field along the constrained trajectory of the system  can then be written as $\mathit{v}_{\gamma(t)}=u^ie_i$. We note that the Poincar\'{e} representation of the geodesic yields the equations for the system dynamics which provides the control inputs required to maintain the system along this trajectory and are given by $\dot{u}^i+{\stackrel{\mathcal{D}}{\Gamma^i}}_{(e)jk} {u}^j{u}^k =0 $, for $i=1,2,...,6$.
Comparing this representation to the kinematic equations of the system for any trajectory $\gamma(t)$ we have: $u^1 = v$, $u^2 = \omega$, $u^3 = v_i$, $u^4 = \omega_i$. 

\subsection{Dynamics of the Cooperative Action}
\label{SubSec: Dynamics of the Cooperative Action}

So far, we have only considered the kinematics of the system and have not yet introduced any dynamics. The Lagrange-d'Alembert's Principle on Riemannian Manifold tells us how to introduce the control forces into the equations of motion. A curve $\gamma(t):[a,b] \to Q$ satisfies the principle for a smooth one-form $F$ as a force and the smooth Lagrangian $L_G$ if and only if 
\begin{equation}
		{\stackrel{G}{\nabla}}_{\mathit{v}_{\gamma(t)}}\mathit{v}_{\gamma(t)}=G^{\#}(F(t,\gamma(t))),
\end{equation}
in which, $G^{\#}:T^{*}Q \to TQ$ is the musical isomorphism associated with the Riemannian metric $G$ \cite{Bloch1996,Bullo2004}.

An extension to the Lagrange-d'Alembert's principle necessitates the  introduction of the input control force to a constrained system described by \eqref{Constrained Dynamics}. This results in following dynamical system equations for the constrained system:
\begin{eqnarray}
	{\stackrel{\mathcal{D}}{\nabla}}_{\mathit{v}_{\gamma(t)}}\mathit{v}_{\gamma(t)} = P(Y(t)),
	\label{eq:Lagrange-dAlembert}
\end{eqnarray}
with $Y(t)=G^{\#}(F(t,\gamma(t)))$. Only the components of the external forces that live in the subspace spanned by the generators of the distribution $\mathcal{D}$ contribute to the motion of the system. The other components of the forces ensure the satisfaction of the constraints. 

Given the unicycle model for each agent considered in this work, the external force consists of the Newtonian force and torque acting on the agents and the load. In geometric terms, the force is a one-form in the vector space of the differential operators, $F=\frac{1}{m}\cos{\theta}fdx+\frac{1}{m}f\sin{\theta}dy+\frac{1}{J}\tau d\theta +\frac{1}{m_i}f_i\cos{\theta_i} dx_i+\frac{1}{m_i}f_i\sin{\theta_i}dy_i+\frac{1}{J_i}\tau_i d\theta_i$.  The constraint force is the projection of $F$ to the subspace spanned by $X_,~X_2,~X_3$, and can be expressed in the chart-induced basis as:
\begin{align}
    P(Y(&t))= h_i \frac{h_if+hf_i}{h^2+h_i^2} \Big( \cos{\theta} \frac{\partial}{\partial x} + \sin{\theta} \frac{\partial}{\partial y} \Big) + \tau \frac{\partial}  {\partial \theta}\nonumber\\
    + & h\frac{h_if+hf_i}{h^2+h_i^2}\Big(\cos{\theta_i}\frac{\partial}{\partial x_i}+\sin{\theta_i}\frac{\partial}{\partial y_i}\Big)
    +\tau_i \frac{\partial}{\partial \theta_i};
    \label{Constrained_Force}
\end{align}
or in the more useful form given by $P(Y(t))=\frac{h_if+hf_i}{h^2+h_i^2}e_1 + \tau e_2 + \frac{h_if+hf_i}{h^2+h_i^2}e_3 + \tau_i e_4$. This generalized force along with the equations describing the system dynamics completes the constrained dynamics. 

\subsection{Controller Synthesis}
\label{SubSec:CONTROLLER SYNTHESIS}

In this work, we assume all vehicles know the position of the load.  We first synthesize a virtual controller for the load to ensure it achieves the desired trajectory.  Then, we synthesis the control inputs of the ASVs that will enable cooperative transport of the load along the computed trajectory. 

To compensate for the inertial effects of the load during transport, we propose a closed-loop control strategy based on the estimate of the load's pose \cite{Hajieghrary2017}. This is achieved by designing an inner control loop to regulate the movement of the load to ensure it asymptotically approaches the reference trajectory. We employ an extended backstepping control design scheme to achieve such a feedback structure \cite{ASMEHajieghrary2016}. 

To derive the control inputs for each ASV from the control input obtained for the load, we note that the first or the third components of the equations in \eqref{eq:Lagrange-dAlembert} provides a feedback structure that can be used to calculate the necessary force the $i^{th}$ vehicle should apply to satisfy the constrained dynamics.  

Conveniently, the torque input for the $i^{th}$ ASV could be designed independently. We leverage this property to design torque inputs that ensures the force applied by the ASVs are feasible and satisfy the system constrains. We accomplish geometrically where vehicles move parallel to the reference trajectory (see Fig. \ref{fig: transport Schematic}) to guarantee the existence of a solution and prevent sudden changes in their linear or angular velocities so as to not require excessive torque or force. We have designed a simple linear controller for the vehicle torque to minimize the attitude error, $e_{\theta}$.      

\section{RESULTS}
\label{Sec:RESULTS}

\begin{figure*}
	\centering
    \begin{subfigure}[b]{0.32\textwidth}
       \centering
       \label{Trajectories}
       \includegraphics[width=1\textwidth]{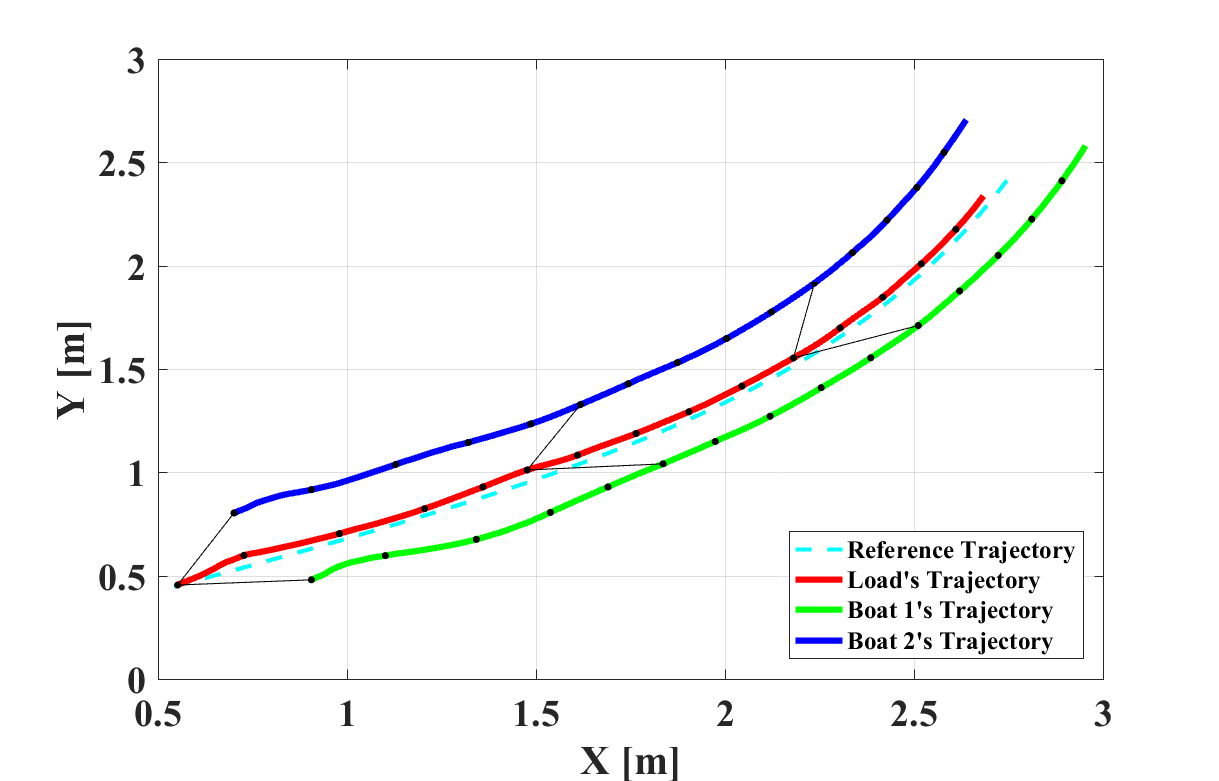}
       \caption{Trajectories}
    \end{subfigure}
   \begin{subfigure}[b]{0.32\textwidth}
       \centering
        \includegraphics[width=1\textwidth]{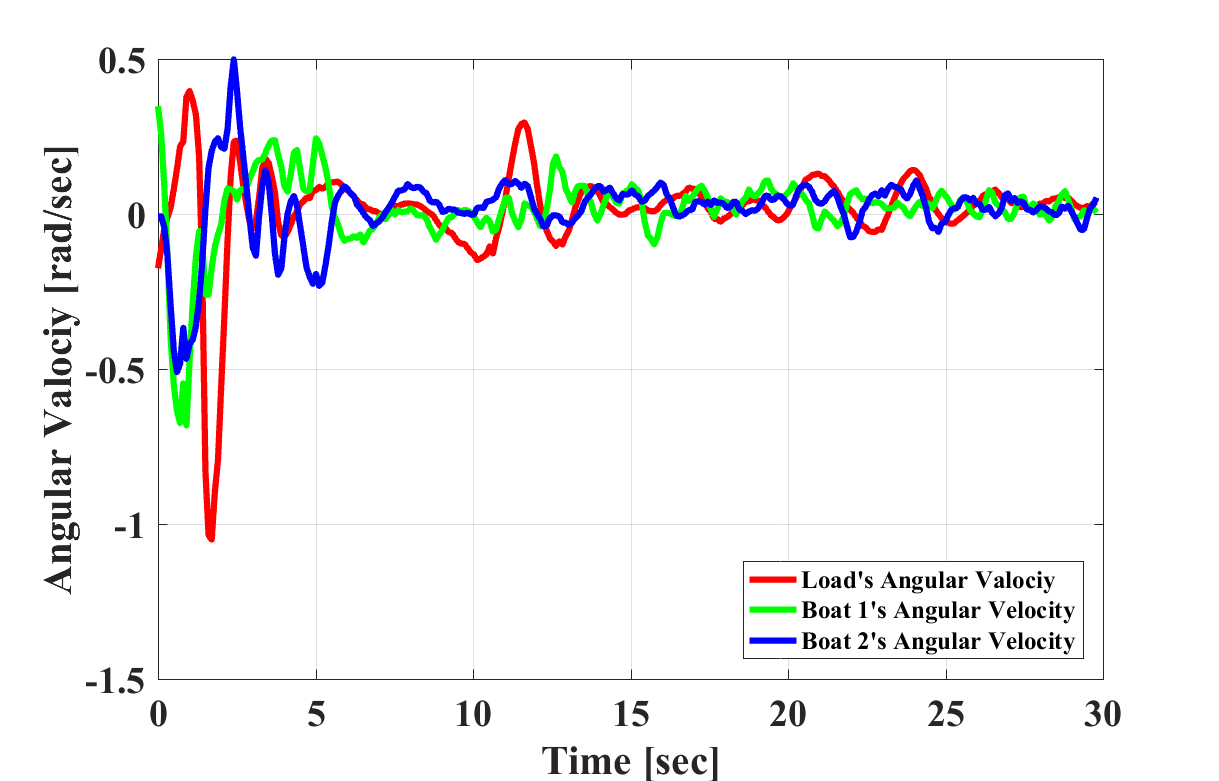}
        \caption{Angular Velocities}
    \end{subfigure}
    \begin{subfigure}[b]{0.32\textwidth}
        \centering
        \includegraphics[width=1\textwidth]{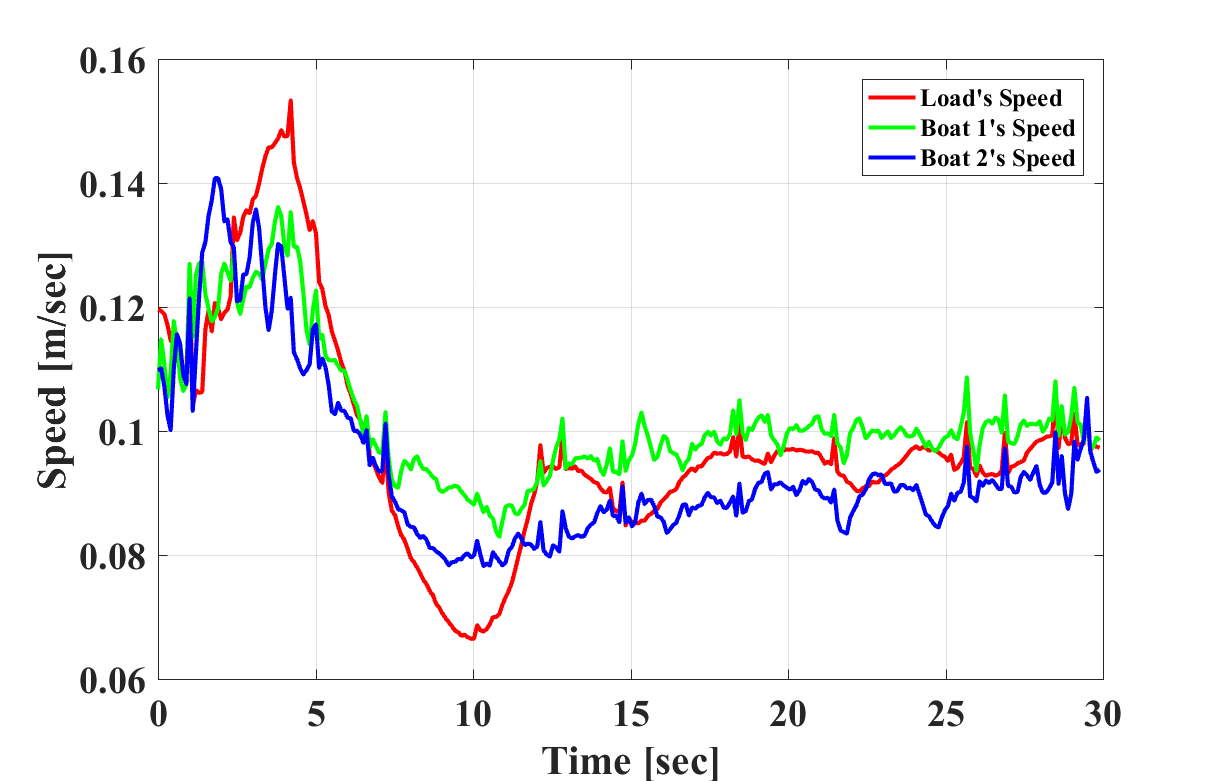}
        \caption{Speed}
    \end{subfigure}\\
    \begin{subfigure}[b]{0.32\textwidth}
       \centering
       \includegraphics[width=1\textwidth]{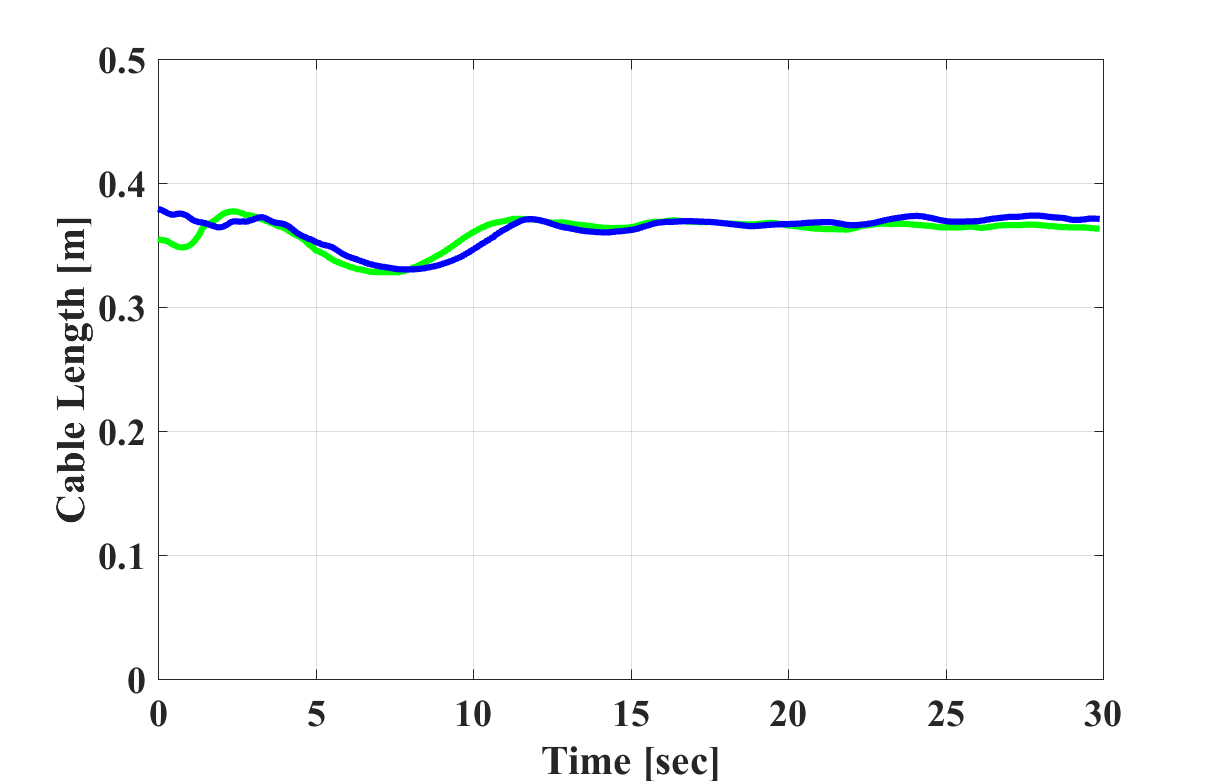}
       \caption{Length of the Connecting Cables}
    \end{subfigure}
   \begin{subfigure}[b]{0.32\textwidth}
       \centering
        \includegraphics[width=1\textwidth]{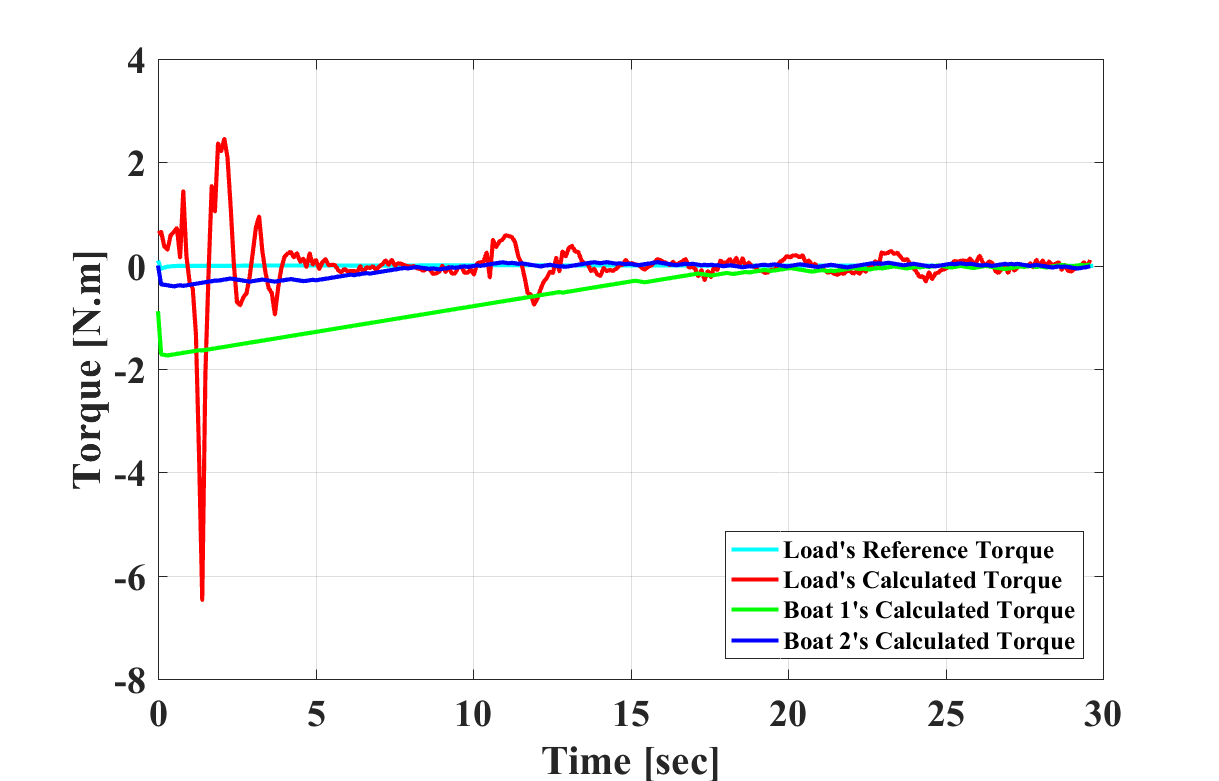}
        \caption{Torques}
    \end{subfigure}
    \begin{subfigure}[b]{0.32\textwidth}
        \centering
        \includegraphics[width=1\textwidth]{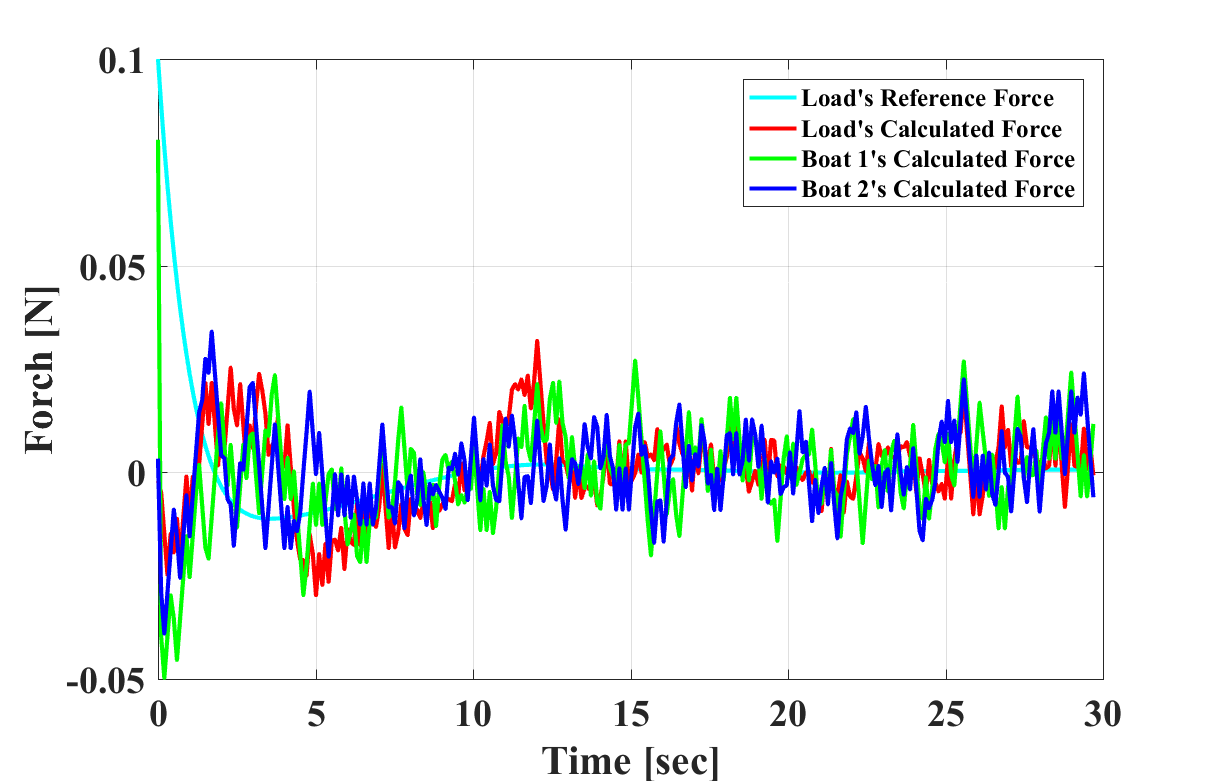}
        \caption{Forces}
    \end{subfigure}
  \caption{The Experimental results for the proposed dynamic \textit{Closed-Loop} control algorithm. In (a) the dotted line indicates the desired position of the load, while the solid line is the position of it read by the motion capture system. At the beginning large torque inputs are applied to the boats to turned them inline with the reference trajectory. During this period the magnitude of the force is high to move the force sufficiently to satisfy the constraints. }
  \vspace{-2mm}
  \label{Experiment Results}
\end{figure*} 

\begin{figure}
	\centering
	\includegraphics[width=0.9\linewidth]{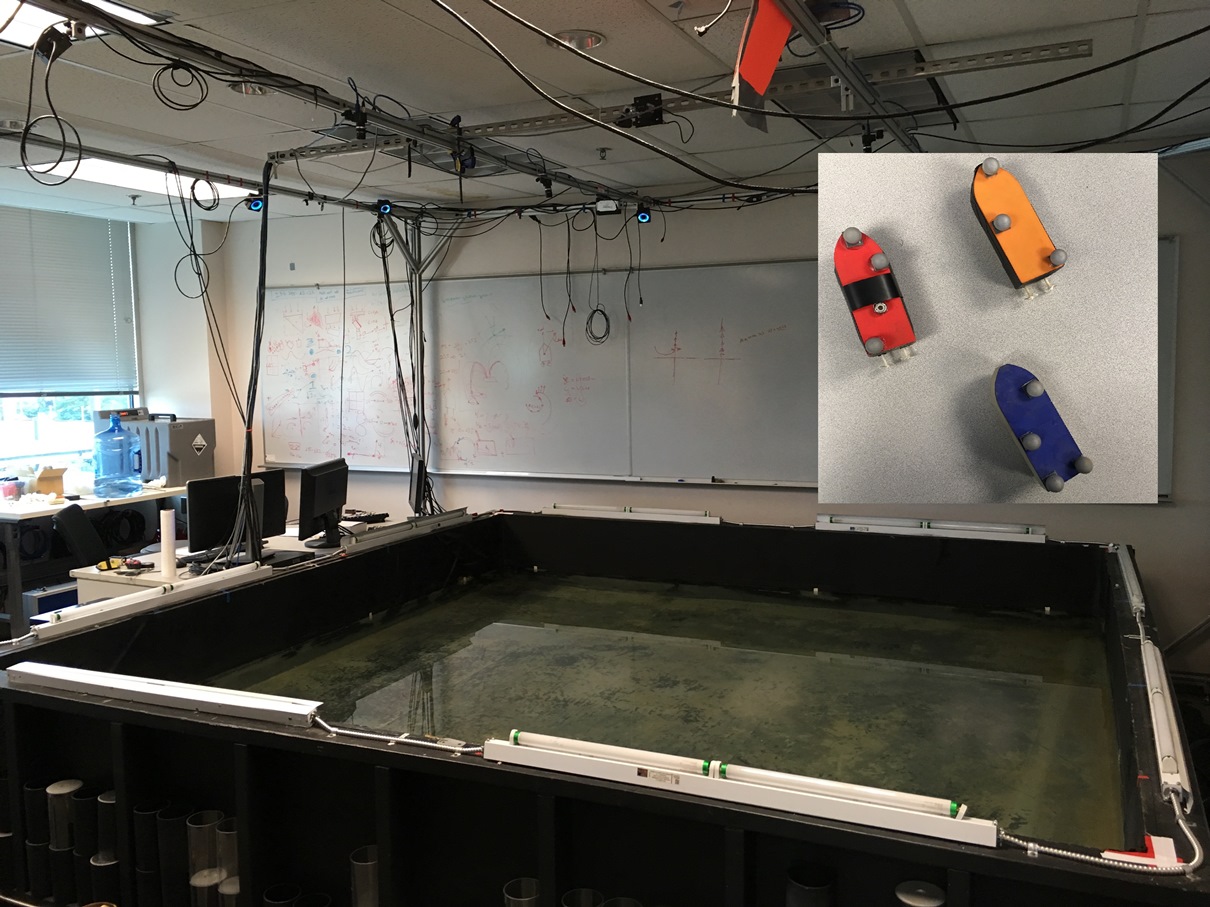}
	\caption{multi-Robot Coherent Structure Testbed (mCoSTe). This is an indoor testbed that consists of a $3m \times 3m \times 1m$ water tank covered with motion capture cameras and a fleet of micro Autonomous Surface Vehicles (mASVs)}
    \vspace{-6mm}
	\label{fig: mCoSTe}
\end{figure}

We validate the proposed control strategy in the multi-Robot Coherent Structure Testbed (mCoSTe), see Fig. \ref{fig: mCoSTe}. The mCoSTe is an indoor laboratory testbed that consists of a $3m \times 3m \times 1m$ water tank and a fleet of micro Autonomous Surface Vehicles (mASVs) \cite{Larkin2014}. The mASVs are differential vehicles equipped with a micro-controller board, a XBee radio module, and an inertial measurement unit (IMU). Each mASV is approximately $12$ $cm$ long with a mass of about $45$ $g$. Localization for the vehicles and load is provided by an external motion capture system.

The objective of the cooperative control is to transport a load such that it tracks the reference trajectory. The chosen reference trajectory in this experiment is given by $y(t) = \tan(\frac{2}{\pi}x(t))$.  The reference trajectory was selected to demonstrate that the proposed strategy can synthesize control inputs to enable the team to maneuver along trajectories with large curvatures while remaining within the physical confines of the tank.  Fig. \ref{Experiment Results} shows the results of the closed-loop experiment. While the initial attitude of the load is far off from the reference trajectory, the closed-loop strategy ensures the ASVs steer the load to settle along the reference trajectory. In the transient, the vehicles must move fast enough to satisfy the constraints on the length of the connecting cables. After the load is maneuvered onto the right track, the force and torque inputs applied by the ASVs reduce significantly.  The oscillations in the bearings and speeds are the result of the low-level speed controllers running onboard the ASVs.  Using larger vehicles would most likely result in much smoother control commands. Videos of more experiments are available at \url{https://www.youtube.com/playlist?list=PLmamVA9vIjfpon8IbsvSVVQO88HqlsOCt}.

By design, the ASVs steer themselves such that they are on trajectory that is parallel to the reference trajectory. Simultaneously, each vehicle's speed is controlled such that the constraint forces move the load on the desired trajectory. The ASV control inputs should ensure that the cable connecting the vehicle to the load remains taut at all times.  In this work, we assume the system initializes in a configuration where the cables are taut. Without this assumption, the vehicles may not succeed in transporting the load along the desired trajectory.  Fig. \ref{Experiment Results}(d) shows how the controller successfully achieves a taut cable configuration when the load drifts closer to the vehicle. When this happens ($\approx 5-10sec$) the controller for the load reduces the load's speed leading in a decrease in the vehicle speeds. These combined effects result in tightening of the cables that connect the load to the vehicles and shows the robustness of the strategy in the presence of disturbances in the load position.  

\section{CONCLUSION}
\label{Sec:CONCLUSION}

In this paper, we addressed the cooperative transport problem involving a team of ASVs towing a buoyant load. We considered the dynamics of the constrained system where each vehicle is attached to the load at the same point.  The approach decomposes the cooperative transport problem into a collection of subsystems each consisting of one ASV and the load whose states evolve on a smooth manifold. Using a differential geometric approach, our strategy integrates the non-holonomic constraints, defined by the system dynamics, and the holonomic constraints, defined by the task requirements, into a single modeling and controller synthesis framework.  We validated the proposed strategy in experiment using a team of ASVs.  Our results show not only feasibility of the strategy but its ability to synthesize controllers that are robust to external disturbances. For future work, we are interested in extending this approach to handle hard constraints on control inputs and develop new strategies to recover from cables losing tension.  


\bibliographystyle{IEEEtran}
\bibliography{Main}

\begin{thebibliography}{10}
\providecommand{\url}[1]{#1}
\csname url@samestyle\endcsname
\providecommand{\newblock}{\relax}
\providecommand{\bibinfo}[2]{#2}
\providecommand{\BIBentrySTDinterwordspacing}{\spaceskip=0pt\relax}
\providecommand{\BIBentryALTinterwordstretchfactor}{4}
\providecommand{\BIBentryALTinterwordspacing}{\spaceskip=\fontdimen2\font plus
\BIBentryALTinterwordstretchfactor\fontdimen3\font minus
  \fontdimen4\font\relax}
\providecommand{\BIBforeignlanguage}[2]{{%
\expandafter\ifx\csname l@#1\endcsname\relax
\typeout{** WARNING: IEEEtran.bst: No hyphenation pattern has been}%
\typeout{** loaded for the language `#1'. Using the pattern for}%
\typeout{** the default language instead.}%
\else
\language=\csname l@#1\endcsname
\fi
#2}}
\providecommand{\BIBdecl}{\relax}
\BIBdecl

\bibitem{Tabuada2005}
P.~Tabuada, G.~J. Pappas, and P.~Lima, ``Motion feasibility of multi-agent
  formations,'' \emph{IEEE Transactions on Robotics}, vol.~21, no.~3, pp.
  387--392, June 2005.

\bibitem{Belta2004}
C.~Belta and V.~Kumar, ``Abstraction and control for groups of robots,''
  \emph{IEEE Transactions on Robotics}, vol.~20, no.~5, pp. 865--875, Oct 2004.

\bibitem{Michael2011}
N.~Michael, J.~Fink, and V.~Kumar, ``\BIBforeignlanguage{English}{Cooperative
  manipulation and transportation with aerial robots},''
  \emph{\BIBforeignlanguage{English}{Autonomous Robots}}, vol.~30, no.~1, pp.
  73--86, 2011.

\bibitem{Wu2014}
G.~Wu and K.~Sreenath, ``Geometric control of multiple quadrotors transporting
  a rigid-body load,'' in \emph{Decision and Control (CDC), 2014 IEEE 53rd
  Annual Conference on}, Dec 2014, pp. 6141--6148.

\bibitem{Lee2014}
T.~Lee, ``Geometric control of multiple quadrotor uavs transporting a
  cable-suspended rigid body,'' in \emph{Decision and Control (CDC), 2014 IEEE
  53rd Annual Conference on}, Dec 2014, pp. 6155--6160.

\bibitem{Cruz2015}
P.~Cruz, M.~Oishi, and R.~Fierro, ``Lift of a cable-suspended load by a
  quadrotor: A hybrid system approach,'' in \emph{American Control Conference
  (ACC), 2015}, July 2015, pp. 1887--1892.

\bibitem{Zeiaee2016}
A.~Zeiaee, R.~Soltani-Zarrin, and R.~Langari, ``A novel approach for tracking
  control of differential drive robots subject to hard input constraints,'' in
  \emph{2016 American Control Conference (ACC)}, July 2016, pp. 2098--2103.

\bibitem{Lewis2000}
A.~D. Lewis, ``Simple mechanical control systems with constraints,'' \emph{IEEE
  Transactions on Automatic Control}, vol.~45, no.~8, pp. 1420--1436, Aug 2000.

\bibitem{Shen2003}
J.~Shen, D.~A. Schneider, and A.~M. Bloch, ``Controllability and motion
  planning of multibody systems with nonholonomic constraints,'' in \emph{42nd
  IEEE International Conference on Decision and Control (IEEE Cat.
  No.03CH37475)}, vol.~5, Dec 2003, pp. 4369--4374 Vol.5.

\bibitem{Cortes2002}
J.~Cortes, S.~Martinez, and F.~Bullo, ``On nonlinear controllability and series
  expansions for lagrangian systems with dissipative forces,'' \emph{IEEE
  Transactions on Automatic Control}, vol.~47, no.~8, pp. 1396--1401, Aug 2002.

\bibitem{Colombo2017}
L.~Colombo, ``A variational-geometric approach for the optimal control of
  nonholonomic systems,'' \emph{International Journal of Dynamics and Control},
  Apr 2017.

\bibitem{Stefani2014}
G.~Stefani, U.~Boscain, J.~Gauthier, A.~Sarychev, and M.~Sigalotti,
  \emph{Geometric Control Theory and Sub-Riemannian Geometry}, ser. Springer
  INdAM Series.\hskip 1em plus 0.5em minus 0.4em\relax Springer International
  Publishing, 2014.

\bibitem{Montgomery2006}
R.~Montgomery, \emph{A Tour of Subriemannian Geometries, Their Geodesics and
  Applications}, ser. Mathematical surveys and monographs.\hskip 1em plus 0.5em
  minus 0.4em\relax American Mathematical Society, 2006.

\bibitem{Bloch2015}
A.~Bloch, L.~Colombo, R.~Gupta, and D.~M. de~Diego, \emph{A Geometric Approach
  to the Optimal Control of Nonholonomic Mechanical Systems}.\hskip 1em plus
  0.5em minus 0.4em\relax Cham: Springer International Publishing, 2015, pp.
  35--64.

\bibitem{Hajieghrary2017}
H.~Hajieghrary, D.~Kularatne, and M.~A. Hsieh, ``Cooperative transport of a
  buoyant load: A differential geometric approach,'' in \emph{2017 IEEE/RSJ
  International Conference on Intelligent Robots and Systems (IROS)}, Sept
  2017, pp. 2158--2163.

\bibitem{Bullo2004}
F.~Bullo and A.~Lewis, \emph{Geometric Control of Mechanical Systems: Modeling,
  Analysis, and Design for Simple Mechanical Control Systems}, ser. Texts in
  Applied Mathematics.\hskip 1em plus 0.5em minus 0.4em\relax Springer New
  York, 2004.

\bibitem{doCarmo2013}
F.~Flaherty and M.~do~Carmo, \emph{Riemannian Geometry}, ser. Mathematics:
  Theory \& Applications.\hskip 1em plus 0.5em minus 0.4em\relax Birkh{\"a}user
  Boston, 2013.

\bibitem{Martinez2007}
E.~Martinez, ``Lie algebroids in classical mechanics and optimal control,''
  \emph{Symmetry, Integrability and Geometry: Methods and Applications
  (SIGMA)}, vol.~3, no. 050, pp. 1--17, Mar 2007.

\bibitem{ACCMaithripala2008}
D.~H.~A. Maithripala, D.~H.~S. Maithripala, and S.~Jayasuriya, ``Unifying
  geometric approach to real-time formation control,'' in \emph{2008 American
  Control Conference}, June 2008, pp. 789--794.

\bibitem{Bloch1996}
A.~M. Bloch, P.~S. Krishnaprasad, J.~E. Marsden, and R.~M. Murray,
  ``Nonholonomic mechanical systems with symmetry,'' \emph{Archive for Rational
  Mechanics and Analysis}, vol. 136, no.~1, pp. 21--99, 1996.

\bibitem{ASMEHajieghrary2016}
H.~Hajieghrary and M.~Ani~Hsieh, ``Dynamic adaptive robust backstepping control
  design for an uncertain linear system,'' \emph{Journal of Dynamic Systems,
  Measurement, and Control}, vol. 138, no.~7, pp. 071\,004--1--071\,004--8,
  2016.

\bibitem{Larkin2014}
D.~Larkin, M.~Michini, A.~Abad, S.~Teleski, and M.~A. Hsieh,
  ``\BIBforeignlanguage{English}{Design of the multi-robot coherent structure
  testbed (mcoste) for distributed tracking of geophysical fluid dynamics},''
  \emph{\BIBforeignlanguage{English}{International Design Engineering Technical
  Conferences and Computers and Information in Engineering Conference}}, vol.
  Volume 5B: 38th Mechanisms and Robotics Conference, p. pp. V05BT08A028, 2014.

\end{thebibliography}

\end{document}